\documentclass[aps,prl,twocolumn,groupedaddress]{revtex4}
\usepackage{graphicx}

\begin{document}


\title{From ballistic transport to tunneling in electromigrated ferromagnetic breakjunctions}


\author{Kirill I. Bolotin}
\author{Ferdinand Kuemmeth}
\author{Abhay N. Pasupathy}
\author{D. C. Ralph}
\email[]{ralph@ccmr.cornell.edu}
\affiliation{Laboratory of Atomic and Solid State Physics, Cornell University, Ithaca, NY 14853}

\date{\today}

\begin{abstract}
We fabricate ferromagnetic nanowires with constrictions whose cross section can be reduced gradually from 100 nm to the atomic scale and eventually to the tunneling regime by means of electromigration. These devices are mechanically stable against magnetostriction and magnetostatic effects. We measure magnetoresistances $\sim$0.3\% for 100 $\times$  30 nm$^2$ constrictions, increasing to a maximum of 80\% for atomic-scale widths. These results are consistent with a geometrically-constrained domain wall trapped at the constriction.  For the devices in the tunneling regime we observe large fluctuations in MR, between -10 and 85\%.
\end{abstract}


\maketitle

The entry of a magnetic domain wall into a nanometer-scale magnetic contact can cause magnetoresistance (MR = [R(AP) - R(P)]/R(P), where R(AP) is the resistance with an antiparallel orientation for the magnetizations in the electrodes and R(P) is the resistance with parallel magnetizations). Several different interesting mechanisms can contribute, depending on the size of the contact.  If the device diameter is greater than tens of nm, the largest contribution to the domain wall resistance is generally the anisotropic MR - a difference in the resistivity of a magnetic material depending on whether the magnetic moment is oriented parallel or perpendicular to the current.  This contribution is relatively small, typically giving MR values of a few percent \cite{eins}.  As the contact diameter is reduced, the width of the domain wall can be constrained by the geometry and decreases in proportion to the contact width \cite{zwei}.  Eventually a new mechanism of MR may become dominant if a domain wall is sufficiently narrow that the spin of a conduction electron cannot follow the direction of the local magnetization adiabatically \cite{drei}. In that case the domain wall can produce increased electron scattering that is analogous to the giant magnetoresistance effect in magnetic multilayers \cite{vier}. For very small metallic contacts, approaching the single-atom diameter regime, values of MR as large as 200\% \cite{fuenf} to 100,000\% \cite{sechs} have been reported, and ascribed to a ``ballistic magnetoresistance'' effect involving scattering of electrons from an atomically-abrupt domain wall.  However, these large effects have not been reliably observed in well-controlled mechanical break junctions \cite{sieben}, and some have argued that the very large changes in resistance are due to the effects of magnetostriction or magnetostatic forces that cause the contact to break and reform as the magnetic field is varied \cite{acht}.  Finally, if the contact diameter is reduced beyond the single-atom limit, it enters the tunneling regime.  MR in that regime reflects the spin polarization of tunneling electrons, and for a nanoscale device it is interesting to ask to what extent the MR depends on the atomic-scale geometry of the contacts, rather than simply the spin polarization of the bulk electron density of states  \cite{neun,zehn}.

In the present work we fabricate two thin-film ferromagnets connected by a small magnetic constriction which can be controllably narrowed by electromigration from about 100 $\times$  30 nm$^2$ to the atomic scale and finally to a tunnel junction. This allows us to study the MR as the contact region between the two ferromagnets is progressively narrowed in a single sample. One advantage of this device geometry is that the magnets are attached rigidly to a non-magnetic substrate with no suspended parts, so that the influence of magnetostriction and magnetostatic forces on the contact are expected to be negligible.  Two previous experiments have also achieved mechanically-stable magnetic devices for MR studies by means of nanopores and ion milling techniques \cite{elf,zwoelf}, but these devices were not tunable to near the atomic scale. 

\begin{figure}
	\includegraphics{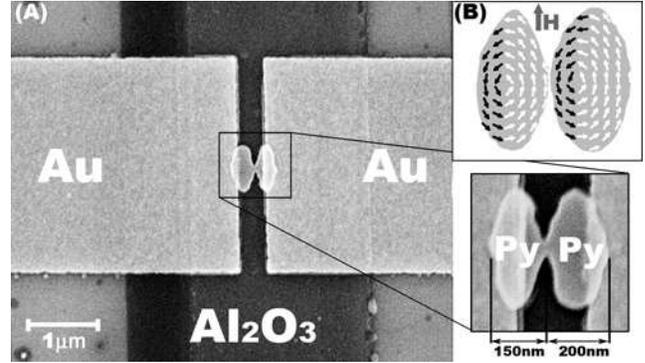}
	\caption{A) Scanning electron micrograph of a finished device.  Gold electrodes are used to contact two permalloy thin-film magnets (inset) on top of an oxidized aluminum gate.  The irregular shape of the Py electrodes results from imperfect liftoff during fabrication. B) Micromagnetic modeling showing antiparallel magnetic alignment across the tunneling gap in an applied magnetic field of H= 66 mT.}
	\label{fig1}
\end{figure}

Our device design builds on the approach taken by Pasupathy \emph{et al.} \cite{dreizehn}, in which the angle between the moments in the two magnetic electrodes could be manipulated by fabricating them with different shapes, so that they undergo magnetic reversal at different values of the applied magnetic field.  However, the shapes of the electrodes used in  \cite{dreizehn} were not optimal in that an accurate antiparallel (AP) configuration of the two moments could not be obtained reliably.  In the design used here (shown in Fig.~\ref{fig1}) the electrodes are elongated along the axis perpendicular to the constriction that connects them.  A magnetic field is applied parallel to the long axis of the electrodes. From simple magnetostatic considerations one expects that dipole interactions between the two electrodes will favor AP alignment with a domain wall in the constriction region for small applied magnetic fields.  For sufficiently strong applied fields, both moments align parallel (P) to the field. We have performed micromagnetic modeling of this geometry using the OOMMF code \cite{vierzehn}. We find that the electrodes can access some states other than the simple uniform P and AP states at intermediate values of field (see the vortex states in Fig.~\ref{fig1}B), but the local magnetizations on opposite sides of the constriction region still accurately remain either P or AP. When the electrode magnetizations are AP and they are connected by a narrow bridge of ferromagnetic metal, a domain wall forms inside the bridge.  We expect that the rotation of the magnetization within the domain wall occurs in plane of the thin-film electrodes, since the demagnetization field of the film will prevent the formation of a Bloch wall. However, the detailed structure of the domain wall is expected to depend on the atomic arrangement in the bridge connecting the two ferromagnets \cite{zehn}.

\begin{figure}
	\includegraphics{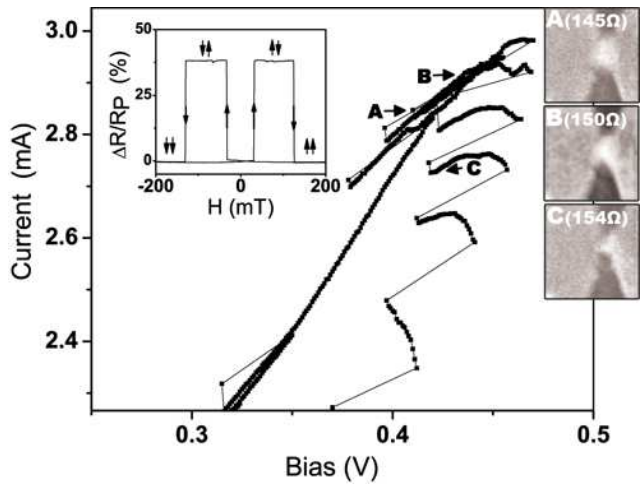}
	\caption{The cross section of a constriction is reduced in stages (A,B,C) by repeatedly ramping the bias voltage until electromigration begins, and then quickly decreasing the bias, following a procedure similar to that in \cite{sechszehn}.  The SEM micrographs illustrate the gradual narrowing of the constriction, which appears bright in these images.  Inset: Resistance as a function of magnetic field for a tunneling device exhibiting abrupt switching between parallel and antiparallel magnetic states. }
	\label{fig2}
\end{figure}

We fabricate the devices on top of an oxidized aluminum gate electrode (not used in this experiment), on a silicon substrate by using aligned steps of electron-beam lithography \cite{fuenfzehn} and thermal evaporation to first deposit gold contact pads 20 nm thick and then the magnetic permalloy electrodes 30 nm thick, with a 100 nm wide permalloy bridge connecting the magnetic electrodes (Fig.~\ref{fig1}A).  We chose permalloy for its low crystalline anisotropy, low magnetostriction and high spin polarization at the Fermi level.

To vary the size of the bridge connecting the two magnetic electrodes, we use controlled electromigration \cite{sechszehn} at liquid helium temperatures.  We slowly ramp the voltage across the constriction while monitoring the current.  At roughly 3 mA (10$^8$ A/cm$^2$), electromigration begins (as indicated by an increase in differential resistance), at which point the acquisition software quickly lowers the bias.  Repeating this procedure allows us to increase the resistance of the junction to any desired value between 100 $\Omega$ and 1 k$\Omega$ with better than 10\% accuracy and to values between 1 k$\Omega$ and 20 k$\Omega$ with better than 50\% accuracy. 

We have imaged the process of controlled electromigration inside a scanning electron microscope using test samples at room temperature. Fig.~\ref{fig2} (A,B,C) shows the gradual narrowing of the constriction as electromigration proceeds, and demonstrates that electromigration produces a single break near the narrow region of the permalloy bridge.  Once the permalloy constriction becomes narrower than about 10 nm, its structure cannot be resolved in the SEM.  Devices for which electromigration is allowed to proceed to form a tunneling gap exhibit magnetic tunnel-junction characteristics (Fig.~\ref{fig2}, inset), with stable switching between well-defined P and AP states.  This agrees well with the micromagnetic simulations.

\begin{figure}
	\includegraphics{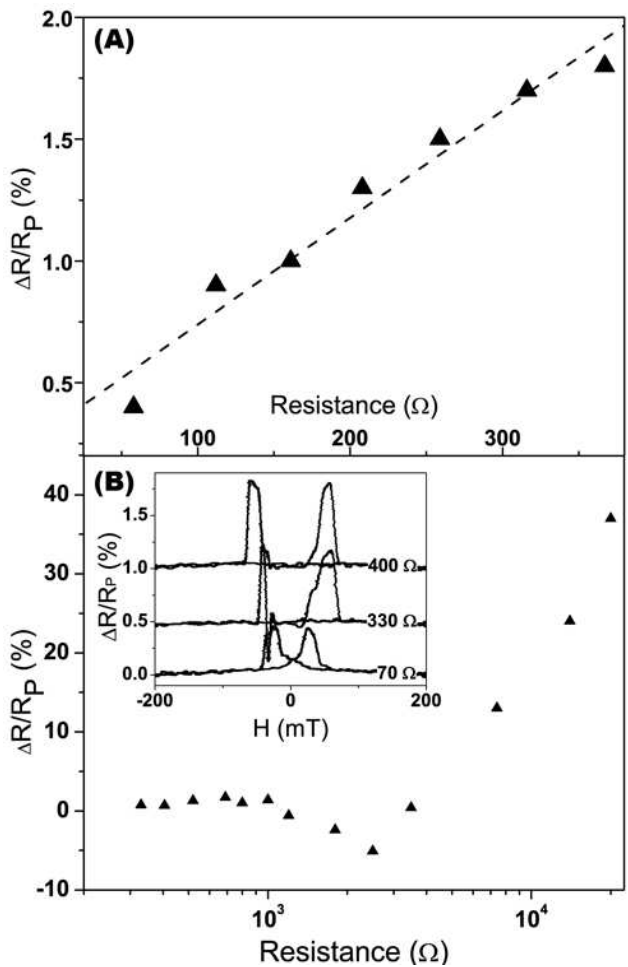}
	\caption{A) Magnetoresistance as a function of resistance in the range less than 400 $\Omega$ (device I).  B) Magnetoresistance as a function of resistance in the range 60 $\Omega$ - 15 k$\Omega$ (device II).  Inset: Switching behavior of device II at different resistances (curves are offset vertically for clarity.)}
	\label{fig3}
\end{figure}

We perform transport measurements at 4.2 K with the samples either immersed in liquid helium or in cryogenic vacuum, to minimize the possibility of oxidation during and after the electromigration process.  Initially, the resistance of each 100 nm wide device is approximately 60~$\Omega$.  The constriction between magnetic electrodes is then progressively narrowed by electromigration, with magnetoresistance measurements made after each stage. When the resistance begins to approach a significant fraction of $h/e^2=25.8~k\Omega$, the transport is likely to be dominated by ballistic transport through just a few apex atoms.  Finally, for devices with resistances of more than approximately $h/e^2$ the transport is dominated by electron tunneling.  We find that the MR properties of the devices are qualitatively different in the regimes of low resistance ($<$ 400 $\Omega$), intermediate resistance (400 $\Omega$ - 25 k$\Omega$) and tunneling ($>$ 25 k$\Omega$), so that we will analyze these regimes separately below.

When the resistance of a device is low ($<$ 400 $\Omega$) it increases smoothly as electromigration proceeds. The cross-section of the constriction varies from 100 $\times$  30 nm$^2$ (60 $\Omega$) to approximately 1 nm$^2$ (400 $\Omega$), with the latter estimate based on the Sharvin formula \cite{siebzehn}.  In this regime we find small ($<$ 3\%) positive MR which increases as the constriction is narrowed (Fig.~\ref{fig3}A). This is consistent with other recent experiments \cite{elf,achtzehn} and well described by the semiclassical theory due to Levy and Zhang \cite{drei}.  In this theory, the resistance of the domain wall scales inversely with its width and the MR ranges typically from 0.7\% to 3\% for bulk ferromagnets.

\begin{figure}
	\includegraphics{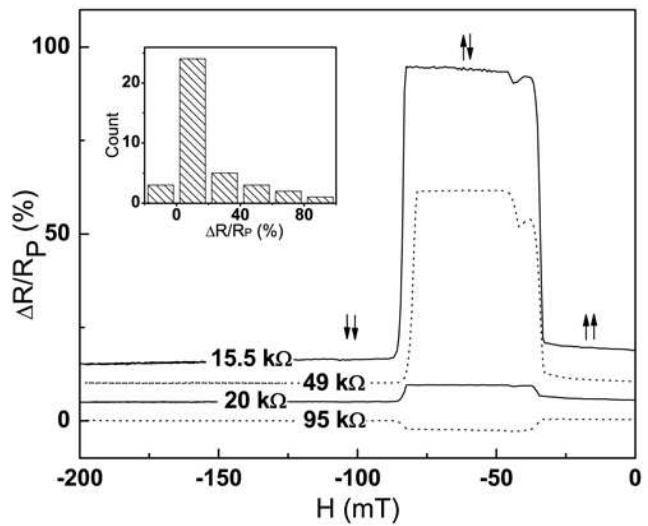}
	\caption{The evolution of magnetoresistance in the tunneling regime (device III) as the resistance of the tunneling gap is changed by electromigration (curves are vertically offset for clarity).  Inset: distribution of magnetoresistances for devices in the tunneling regime ($>$ 20 k$\Omega$).}
	\label{fig4}
\end{figure}

The resistance range from 400 $\Omega$ to 25 k$\Omega$ corresponds to a crossover between ballistic transport through just a few atoms and tunneling.  In this regime the resistance of the device increases in discrete steps during electromigration and the process is less controllable.  Similar behavior is seen in conventional mechanical break junctions and corresponds to the rearrangement of atoms in the constriction \cite{neunzehn}.  In this intermediate regime, the value of MR exhibits pronounced dependence on the resistance of the device.  The MR has a minimum for resistances above 1 k$\Omega$, and typically changes sign here to give negative values.  As the resistance is increased further into the k$\Omega$ range, the MR increases gradually to positive values of 10-20\%.  These trends are reproducible, although the exact dependence of the MR on resistance differs from device to device.  The highest MR that we have observed was 80\% for a device with a resistance of 14.5 k$\Omega$.  The MR values that we observe in the point contact regime are smaller than expected from scaling results of the semiclassical theory \cite{drei}.  This difference is not surprising when the current is transmitted through just a few quantum channels \cite{zehn}.
The MR curves for all of the samples in the metallic regimes, with resistances below $h/e^2$, do not exhibit abrupt transitions between the P and AP resistances when the magnetic field is swept, but rather show more gradual and irregular behavior (Fig.~\ref{fig3}B, inset).  The form of the MR curve also varies as a device's resistance is increased.  This suggests that the position and structure of the domain wall in the constriction may change as the magnetic field is varied.
 
When the resistance of a device becomes greater than tens of k$\Omega$, the transport is dominated by electron tunneling. In this regime most devices exhibit clean switching behavior with well defined P and AP states (Fig.~\ref{fig2}, inset).  Even after the metal bridge is broken, the size of the tunneling gap can still be adjusted by further electromigration.  The shape of the MR curve and the values of the switching fields do not change significantly as electromigration changes the tunnel gap, but the value of the MR and even its sign can fluctuate over a wide range (Fig.~\ref{fig4}).  This suggests that the MR is sensitive to the details of the atomic structure near the tunnel gap.  The tunneling current is flowing through just a few atoms on each of the electrodes, and the electronic structure at these atoms does not necessarily reflect the same degree of spin polarization as in the bulk of the ferromagnet \cite{zehn}.  The histogram of MR values we measure over 38 values of resistance in 20 tunneling devices is shown in the inset of Fig.~\ref{fig4}.  The MR values range from -10\% to a maximum of 85\%.

\begin{figure}
	\includegraphics{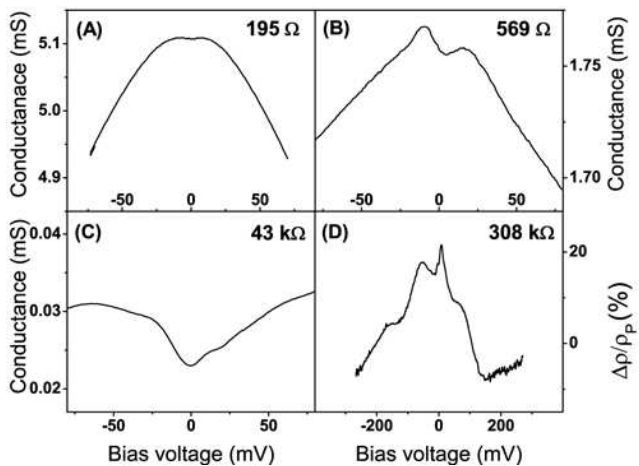}
	\caption{Differential conductance as a function of bias voltage for device IV at small (A), intermediate (B) and large (C) resistances.  D) Magnetoresistance as a function of bias for a tunneling device (device V).}
	\label{fig5}
\end{figure}

It is interesting to examine the bias voltage (\emph{V}) dependence of the conductance and the MR. In the low resistance regime ($<$ 400 $\Omega$), the MR is independent of \emph{V} and the conductance decreases with bias (Fig.~\ref{fig5}A), as it should because of increased backscattering at higher \emph{V} in metallic devices. For high-resistance tunneling devices the conductance increases with \emph{V} (Fig.~\ref{fig5}C).  In the intermediate regime (400 $\Omega$ to 25 k$\Omega$), the conductance typically increases with \emph{V} near \emph{V}=0 as in the tunneling regime, but then decreases with \emph{V} at higher biases as in the metallic regime.  We interpret this as the effect of having both metallic channels and tunneling channels contributing in parallel. The MR in the tunneling regime displays strong dependence on \emph{V} (Fig.~\ref{fig5}D). The exact form of the dependence differs from device to device, but typically the MR drops by a factor of 2 on the scale of \emph{V} = 100 mV.  This is similar to the behavior of standard magnetic tunnel junctions with oxide barriers \cite{neun} and is in contrast to the STM experiments by Wulfhekel et al. \cite{zwanzig}, in which no voltage dependence of the MR was found for a vacuum tunneling gap.

In summary, using a combination of electron beam lithography and controlled electromigration, we fabricate ferromagnetic junctions with tunable cross section, with sizes ranging from 100 $\times$  30 nm$^2$ to near the atomic scale. Further electromigration opens a tunneling gap between the electrodes. These devices do not have any suspended parts and are stable against magnetostriction and magnetostatic effects. We measure the magnetoresistance as a function of the cross-section of the constriction. When the cross-section is larger than approximately 1 nm$^2$ the MR is less than 3\% and increases as the cross-section decreases. For near-atomic-sized constrictions we observe MR as high as 80\%, but find no devices in which the MR is as large as reported previously for the ballistic magnetoresistance mechanism \cite{fuenf,sechs}.  In the tunneling regime the MR values fluctuate over a wide range, -10\% to 85\%, even for small changes in the atomic structure near the constriction in a single device. 

We note that Keane, Lu and Natelson have recently posted independent results of a similar experiment \cite{einundzwanzig}.

\begin{acknowledgments}
We thank O. Ozatay and A. Champagne for useful discussions and V. Sazonova for experimental help.  This work was funded by the NSF (DMR-0244713 and through the use of the Cornell Nanoscale Facility/NNIN) and by the ARO (DAAD19-01-1-0541).

\end{acknowledgments}


\end{document}